\documentclass[superscriptaddress,aps,prd,nofootinbib,twocolumn,showpacs]{revtex4-1}

\usepackage{multirow}
\usepackage{amsmath,epsfig,ulem}
\usepackage{amsfonts}
\usepackage{amsmath}
\usepackage{amssymb}
\usepackage{graphicx}
\usepackage{mathrsfs}
\usepackage{hhline,color}
\usepackage{pifont}
\usepackage{lmodern}
\usepackage[utf8]{inputenc}

\def\beq{\begin{equation}}
\def\eeq{\end{equation}}
\def\beqa{\begin{eqnarray}}
\def\eeqa{\end{eqnarray}}
\def\ban{\begin{eqnarray*}}
\def\ean{\end{eqnarray*}} 
\def\bi{\begin{itemize}}
\def\ei{\end{itemize}}

\newcommand{\Z}{\mathbb{Z}}

\begin{document}

\title{Phase transition and critical end point driven by an external magnetic field in asymmetric
quark matter}
\author{Pedro Costa}
\affiliation{Centro de F\'{\i}sica Computacional, Department of Physics,
University of Coimbra, P-3004 - 516  Coimbra, Portugal}
\author{M\'arcio Ferreira}
\affiliation{Centro de F\'{\i}sica Computacional, Department of Physics,
University of Coimbra, P-3004 - 516  Coimbra, Portugal}
\author{Hubert  Hansen}
\affiliation{IPNL, Universit\'{e} de Lyon/Universit\'{e} Lyon 1, CNRS/IN2P3,
4 rue E.Fermi, F-69622 Villeurbanne Cedex, France}
\author{D\'{e}bora P. Menezes}
\affiliation{Departamento de F\'{\i}sica, CFM, Universidade Federal de Santa
Catarina, Florian\'opolis, SC, CP. 476, CEP 88.040 - 900 - Brazil}
\author{Constan\c ca Provid\^encia}
\affiliation{Centro de F\'{\i}sica Computacional, Department of Physics,
University of Coimbra, P-3004 - 516  Coimbra, Portugal}

\date{\today}

\begin{abstract}
The location of the critical end point (CEP) in the QCD phase diagram is
determined under  different scenarios. The effect of strangeness,
isospin/charge asymmetry and an external magnetic field is investigated.
The discussion is performed within the 2+1 flavor
Nambu--Jona-Lasinio model with Polyakov loop.
It is shown that isospin asymmetry shifts the CEP to larger baryonic chemical
potentials and smaller temperatures. At large  asymmetries the CEP disappears.
However, a strong enough magnetic field drives the system into a first order
phase transition.

\end{abstract}


\maketitle

\section{Introduction}

Presently the study of the phase diagram of QCD is the subject of both
theoretical and experimental studies under extreme conditions of 
density and temperature. In particular, it is expected that the
phenomenon of deconfinement occurs in relativistic heavy-ion 
collisions and in the interior of compact stars, two very different 
scenarios when isospin asymmetry is considered. 
While in heavy ion collisions (HIC) the proton fraction is presently not smaller 
than $\sim0.4$, much smaller proton fractions are expected in the  interior 
of neutron stars. The effect of isospin/charge asymmetry in the QCD phase 
diagram has recently been discussed  in \cite{ueda13} and it was shown that 
for a sufficiently asymmetric system the critical end point (CEP) is not 
present \cite{ueda13,abuki13}.

Another degree of freedom that must be considered when discussing
the QCD phase diagram  is strangeness. 
In the interior of a neutron star it is expected that strangeness is 
present in the form of hyperons, of a kaon condensate or of a 
core of deconfined quark matter. 
$\beta-$equilibrium is energetically favored and the Fermi pressure of 
neutrons is reduced if strangeness degrees of freedom are generated
through the action of the weak interaction. On the other hand, the
strong force  governs  heavy ion collisions.

The  compact astrophysical objects known as magnetars \cite{duncan}, 
which include soft gamma repeaters and anomalous x-ray pulsars, 
are expected to have very strong magnetic fields in their interior. 
Extremely strong magnetic fields are also expected to affect the 
measurements in heavy ion collisions at very high energies \cite{HIC} 
or the behavior of the first phases of the Universe \cite{cosmo}. 
Fields of this intensity affect the QCD phase diagram as shown in 
\cite{avancini2012}.
Therefore, understanding the effect of an external magnetic field 
on the structure of the QCD phase diagram is very important, 
and this has already led to several studies
\cite{Klimenko:1991he,Ebert:2003yk,Ferrer:2005vd,Mizher:2010zb,Chatterjee:2011ry,
Chernodub:2011mc,Ferreira:2013oda}, in particular, at zero chemical potential $\mu=0$ 
(the $T-eB$ plane); see \cite{baliJHEP2012,ruggieri2,Fraga:2012rr,DElia3} for a review. 

At zero chemical potential, almost all low-energy effective models, 
including the Nambu-Jona-Lasinio (NJL)-type models, as well as some lattice QCD (LQCD) calculations
\cite{D'Elia:2010nq,D'Elia:2011zu,Braguta:2010ej,Ilgenfritz:2012fw,Ilgenfritz:2013ara}, 
found an enhancement of the condensate due to the magnetic field (magnetic catalysis) 
independently of the temperature. 
However, more recently, LQCD studies \cite{baliJHEP2012,Bali:2012zg}, for $N_f=2+1$ flavors 
with physical quarks and pion masses, show a suppression of the light condensates 
by the magnetic field in the transition temperature region, an effect known as inverse 
magnetic catalysis \cite{Chao:2013qpa,Fukushima:2012kc,Kojo:2012js}. 
Indeed, near the transition temperatures the condensate shows a nonmonotonic 
behavior decreasing with $eB$. 
Also interesting is the fact that new lattice QCD calculations report a rise of the 
Polyakov loop with $eB$ at the pseudocritical temperature and $eB\lesssim0.8$ GeV$^2$ 
indicating an inverse magnetic catalysis \cite{Ilgenfritz:2013ara}. However, as
pointed out by the authors at a sufficiently strong magnetic field strength the magnetic 
catalysis is seen to be in agreement with almost all effective models that predict magnetic
catalysis at any temperature and magnetic field strength, like the Nambu--Jona-Lasinio model 
with Polyakov loop (PNJL).

In \cite{Ferreira:2013tba}, it has been shown that within the entangled
Nambu--Jona-Lasinio model with Polyakov loop (EPNJL) \cite{sasaki10} the inverse magnetic 
catalysis at $\mu=0$ could be reproduced with a magnetic field dependent parameter $T_0(eB)$ 
in the Polyakov loop. The magnetic field dependence of this parameter mimics the reaction 
of the gluon sector to the presence of an external magnetic field.

The inverse magnetic catalysis mechanism does not occur only at $\mu=0$ and large 
temperatures. This phenomenon has also been obtained at finite chemical potential and
zero or low temperatures: the critical chemical potential for the phase transition 
decreases with  increasing $eB$. This is, however, a region of the QCD phase diagram 
still not accessible to LQCD. In the NJL model the first studies were performed 
in Ref. \cite{Ebert} at $T=0$ and in Ref. \cite{Inagaki} for the full $T-\mu-B$ case. 
This effect has also been obtained in other models \cite{avancini2012,ruggieri14} 
and is the result of a competition between the decrease of the free energy due to the 
condensation in the magnetic field and the increase of the free energy due to the accommodation 
of more valence quarks in the phase space \cite{ruggieri14}. 
In the present work, the same effect will  be obtained. 
In this context, QCD-like models are very useful in the region of moderate temperature 
and chemical potential in the presence of an external magnetic field. 

In the present work we investigate several scenarios of interest for the study of 
either heavy ion collisions or compact stars. 
We show how the CEP changes with the isospin asymmetry and confirm previous results obtained 
within other models that indicate that at sufficiently high asymmetry it does not exist
\cite{ueda11,abuki13,ueda13}. 
We also consider the effect of strangeness in the QCD phase diagram by analyzing different 
chemical equilibrium conditions. 
Finally, we calculate the effect of an external magnetic field on the same scenarios 
previously discussed for a nonmagnetized system.
It will be shown that the magnetic field, if sufficiently strong can drive a first order 
phase transition in an isospin asymmetric system at a quite low temperature. 
The discussion is performed  within the 2+1 flavor PNJL \cite{PNJL}, and 
for reference some results calculated within the NJL model are also included.

\section{Model and Formalism}
\label{sec:model}

Most of the properties of the quark condensates in the presence of an 
external magnetic field were previously obtained with the two flavor 
version of the PNJL and EPNJL models \cite{ruggieri,ruggieri2}. 
Recently, the effects of an external magnetic field on the fluctuations and 
correlations of the quark number and conserved charges, were studied in the 2+1 
PNJL model \cite{Fu}. 

In the present work we describe the quark matter subject to strong magnetic fields 
within the 2+1 PNJL model. 
The PNJL Lagrangian with explicit chiral symmetry breaking where 
the quarks couple to a (spatially constant) temporal background gauge field, 
represented in terms of the Polyakov loop and in the presence of an external 
magnetic field is given by \cite{PNJL}
\begin{eqnarray}
{\cal L} &=& {\bar{q}} \left[i\gamma_\mu D^{\mu}-
	{\hat m}_f \right ] q ~+~ {\cal L}_{sym}~+~{\cal L}_{det} \nonumber\\
&+& \mathcal{U}\left(\Phi,\bar\Phi;T\right) - \frac{1}{4}F_{\mu \nu}F^{\mu \nu},
	\label{Pnjl}
\end{eqnarray}
where the quark sector is described by the  SU(3) version of the
NJL model which includes scalar-pseudoscalar (chiral invariant) 
and the t'Hooft six fermion interactions (that models the axial $U(1)_A$ 
symmetry breaking) \cite{njlsu3}, with ${\cal L}_{sym}$ and ${\cal L}_{det}$ 
given by \cite{Buballa:2003qv}
\begin{eqnarray*}
	{\cal L}_{sym}= G \sum_{a=0}^8 \left [({\bar q} \lambda_ a q)^2 + 
	({\bar q} i\gamma_5 \lambda_a q)^2 \right ] ,
\end{eqnarray*}
\begin{eqnarray*}
	{\cal L}_{det}=-K\left\{{\rm det} \left [{\bar q}(1+\gamma_5) q \right] + 
	{\rm det}\left [{\bar q}(1-\gamma_5)q\right] \right \}
\end{eqnarray*}
where $q = (u,d,s)^T$ represents a quark field with three flavors, 
${\hat m}_f= {\rm diag}_f (m_u^0,m_d^0,m_s^0)$ is the corresponding (current) mass matrix,
{ $\lambda_0=\sqrt{2/3}I$  where $I$ is the unit matrix in the three flavor space, 
and $0<\lambda_a\le 8$ denote the Gell-Mann matrices.
The coupling between the (electro)magnetic field $B$ and quarks, and between the 
effective gluon field and quarks is implemented  {\it via} the covariant derivative 
$D^{\mu}=\partial^\mu - i q_f A_{EM}^{\mu}-i A^\mu$ 
where $q_f$ represents the quark electric charge ($q_d = q_s = -q_u/2
= -e/3$),  $A^{EM}_\mu$ and 
$F_{\mu \nu }=\partial_{\mu }A^{EM}_{\nu }-\partial _{\nu }A^{EM}_{\mu }$ 
are used to account for the external magnetic field and 
$A^\mu(x) = g_{strong} {\cal A}^\mu_a(x)\frac{\lambda_a}{2}$ where
${\cal A}^\mu_a$ is the SU$_c(3)$ gauge field.
We consider a  static and constant magnetic field in the $z$ direction, 
$A^{EM}_\mu=\delta_{\mu 2} x_1 B$.
In the Polyakov gauge and at finite temperature the spatial components of the 
gluon field are neglected: 
$A^\mu = \delta^{\mu}_{0}A^0 = - i \delta^{\mu}_{4}A^4$. 
The trace of the Polyakov line defined by
$ \Phi = \frac 1 {N_c} {\langle\langle \mathcal{P}\exp i\int_{0}^{\beta}d\tau\,
A_4\left(\vec{x},\tau\right)\ \rangle\rangle}_\beta$
is the Polyakov loop which is the {\it exact} order parameter of the $\Z_3$ 
symmetric/broken phase transition in pure gauge.

To describe the pure gauge sector an effective potential 
$\mathcal{U}\left(\Phi,\bar\Phi;T\right)$ is chosen in order to reproduce 
the results obtained in lattice calculations \cite{Ratti:2006},
\begin{eqnarray}
	& &\frac{\mathcal{U}\left(\Phi,\bar\Phi;T\right)}{T^4}
	= -\frac{a\left(T\right)}{2}\bar\Phi \Phi \nonumber\\
	& &
	+\, b(T)\mbox{ln}\left[1-6\bar\Phi \Phi+4(\bar\Phi^3+ \Phi^3)-3(\bar\Phi \Phi)^2\right],
	\label{Ueff}
\end{eqnarray}
where $a\left(T\right)=a_0+a_1\left(\frac{T_0}{T}\right)+a_2\left(\frac{T_0}{T}\right)^2$, 
$b(T)=b_3\left(\frac{T_0}{T}\right)^3$.
The standard choice of the parameters for the effective potential $\mathcal{U}$ is
$a_0 = 3.51$, $a_1 = -2.47$, $a_2 = 15.2$, and $b_3 = -1.75$.

The parameter $T_0$ is the critical temperature for the deconfinement 
phase transition within a pure gauge approach: it was fixed to a constant $T_0=270$ MeV, 
according to lattice findings. Different criteria for fixing $T_0$ may be found in the 
literature, and one of them takes into account the quark backreaction effects on the 
Polyakov loop \cite{wambach}. One should notice, however, that the 
behavior of the relevant physical quantities remains qualitatively the same.

The model being an effective one (up to the scale $\Lambda_{QCD}$) and not renormalizable, 
we use as a regularization scheme a sharp cutoff, $\Lambda$, in 3-momentum space, 
only for the divergent ultraviolet integrals. 
The parameters of the model, $\Lambda$, the coupling constants $G$ and $K$
and the current quark masses $m_u^0$ and $m_s^0$ are determined  by fitting
$f_\pi$, $m_\pi$ , $m_K$, and $m_{\eta'}$ to their experimental values in vacuum. 
Besides, the fifth quantity needed to adjust the parameters of the NJL sector
of the model is an estimation of the quark condensate in the vacuum. 
We consider $\Lambda = 602.3 \, {\rm MeV}$ , $m_u^0= m_d^0=\,  5.5 \,{\rm MeV}$,
$m_s^0=\,  140.7\, {\rm MeV}$, $G \Lambda^2= 1.385$ and $K \Lambda^5=12.36$
as in \cite{Klev_param}.

The thermodynamical potential for the three flavor quark sector, $\Omega$, 
in the mean field approximation is written as
\begin{eqnarray}
	\Omega(T,\,B)&=& 2G \sum_{f=u,\,d,\,s} \left\langle \bar{q}_fq_f\right\rangle^2 
	-4K \, \left\langle \bar{q}_uq_u\right\rangle \left\langle \bar{q}_dq_d\right\rangle
	\left\langle \bar{q}_sq_s\right\rangle  \nonumber\\
	&+&\left(\Omega_f^{vac}+\Omega_f^{mag}+\Omega_f^{med}\right),
	\label{Omega}
\end{eqnarray}
where the vacuum $\Omega_f^{vac}$, the magnetic $\Omega_f^{mag}$, 
the medium contributions $\Omega_f^{med}$ and the quark condensates 
$\left\langle \bar{q}_fq_f\right\rangle$ have been evaluated with 
great detail in \cite{prc,hotnjl}. 

To obtain the mean field equations we must minimize the thermodynamical 
potential (\ref{Omega}) with respect to the order parameters
$\left\langle \bar{q}_fq_f\right\rangle$, $\Phi$ and $\bar{\Phi}$.

\section{The CEP}
\label{sec:cep}

\begin{figure*}[tb]
	\includegraphics[width=0.475\linewidth,angle=0]{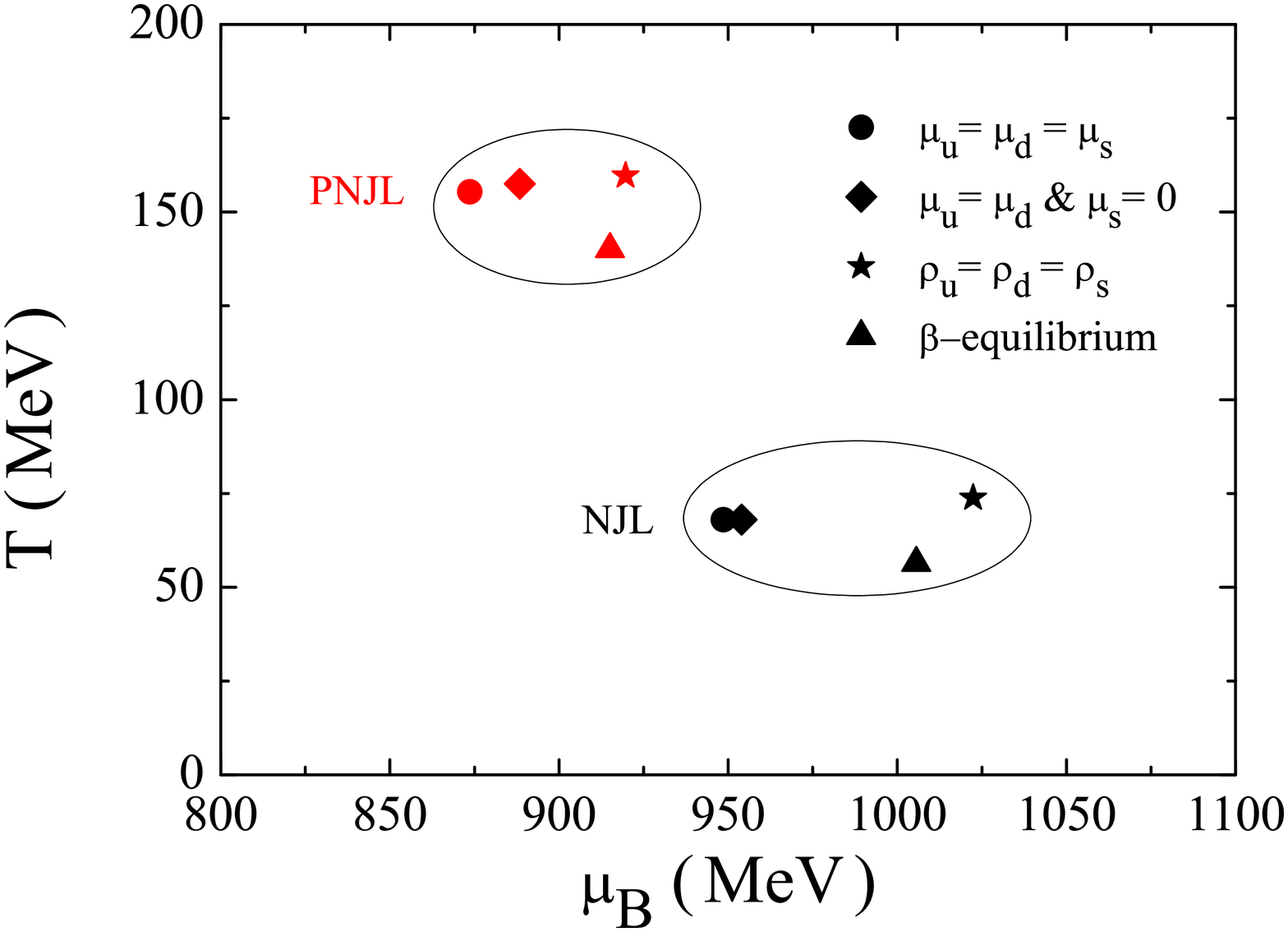} 
	\includegraphics[width=0.475\linewidth,angle=0]{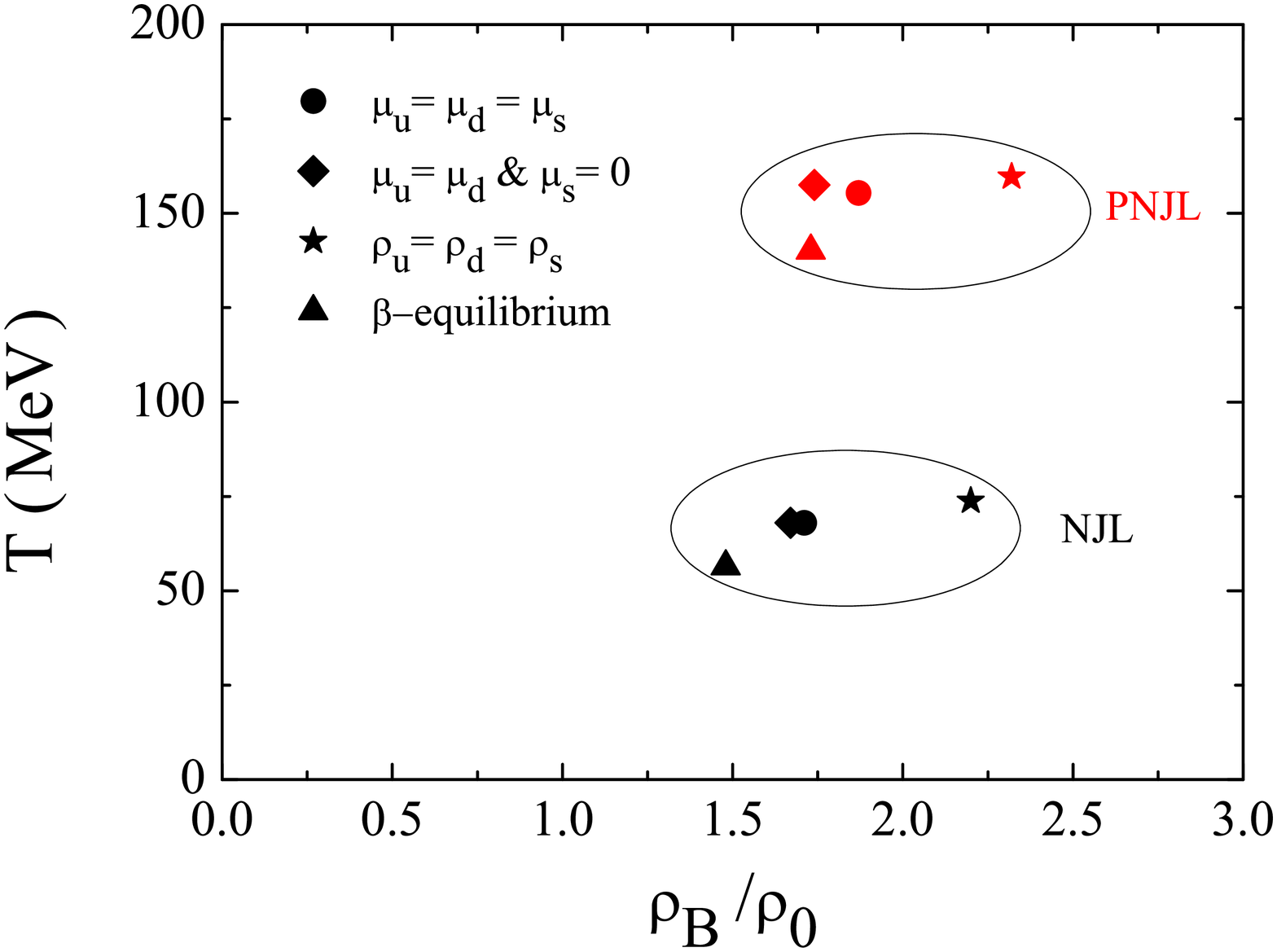}
	\caption{Location of the CEP on a diagram $T$ versus the baryonic
	chemical potential (left) and $T$ versus the baryonic density (right) 
	under different scenarios [all quark chemical potentials equal (circle), 
	the strange quark chemical potential equal to zero (diamond), all quark 
	densities equal (star), and $\beta-$equilibrium condition (triangle)] and
	models (NJL, PNJL). No external magnetic field is considered.} 
	\label{dfase1}
\end{figure*}

In the present section different scenarios obtained by choosing different 
values of the isospin and the strangeness chemical potentials are discussed. 
In terms of the baryon,  charge and strangeness chemical potentials, 
$\mu_B,\,\mu_{Q},\ \mu_S$ we have
\begin{eqnarray}
\mu_{u}&=&\frac{1}{3}\mu_{B}+\frac{2}{3}\mu_{Q},\quad
\mu_{d} = \frac{1}{3}\mu_{B}-\frac{1}{3}\mu_{Q},	\nonumber\\
&&\textrm{and}\quad
\mu_{s}=\frac{1}{3}\mu_{B}-\frac{1}{3}\mu_{Q}-\mu_{S}. \label{chem}
\end{eqnarray}

{\it No external magnetic field.---}We first investigate the location of the 
CEP when no external magnetic field is present. 
We consider the  models NJL and PNJL and the following  different scenarios: 
\textbf{(a)}  Equal quark chemical potentials as done in most calculations. 
This scenario corresponds to zero charge (or isospin) chemical potential
and  zero strangeness chemical potential ($\mu_Q=\mu_S=0$). 
\textbf{(b)} equal $u$ and $d$ quark chemical potentials and a zero strange quark 
chemical potential, corresponding to zero charge (isospin) chemical potential 
and a strangeness chemical potential equal to one-third of the total baryonic 
chemical potential ($\mu_Q=0;\,\mu_S=1/3\mu_B$).
\textbf{(c)}
symmetric matter with equal amounts of quarks $u$, $d$, and $s$, known as 
strange matter ($\rho_u=\rho_d=\rho_s$).
\textbf{(d)} $\beta-$equilibrium matter corresponding to $\mu_u-\mu_d= \mu_{Q}=-\mu_e$ 
and $\mu_d=\mu_s$ ($\mu_S=0$). The temperature, baryonic chemical potential and density 
of the CEPs are given in Table \ref{table:CEP}.

We next analyze Fig. \ref{dfase1} to compare the different scenarios.
The NJL results are shown just for reference.  As already discussed in 
\cite{costa}, the Polyakov loop shifts the CEP to higher temperatures and 
slightly smaller baryonic chemical potentials. 
Matter with the largest isospin asymmetry in this figure is represented by the 
$\beta-$equilibrium results. The $\beta-$equilibrium CEP occurs for one of the 
largest chemical potentials, only slightly below the one obtained for strange matter.
However, it is interesting to see that  for $\beta-$equilibrium the CEP comes 
at lower temperatures.
The reason becomes clear when analyzing the right panel of Fig. \ref{dfase1}: 
the $\beta-$equilibrium matter being less symmetric is less bound and, 
therefore, the transition to a chirally symmetric phase occurs at a smaller 
temperature and density than the symmetric case.

\begin{table}[t]
\begin{center}
    \begin{tabular}{|c||c|c|c||c|c|c|}
            \hline
            \multicolumn{0}{|c||}{$ $}
                & \multicolumn{3}{|c||}{NJL}
                & \multicolumn{3}{|c|}{PNJL} \\
            \hline
           	& $T$ & $\mu_B$ & $\rho_B/\rho_0$ & $T$ & $\mu_B$ & $\rho_B/\rho_0$ \\
        		& [MeV]     & [MeV]     & { } & [MeV] & [MeV] &   { }   \\
        \hline\hline
        $\mu_u=\mu_d=\mu_s$   						& 68	& 948.5	 	& 1.71 & 155.4 	& 873.8	& 1.87             \\
        \hline
        $\mu_u=\mu_d\textmd{; }\mu_s=0$ & 68 	& 954.0	 	& 1.67 & 157.5  & 890.4 &  1.73             \\
        \hline
        $\rho_u=\rho_d=\rho_s$ 						& 73.8 & 1022.4		& 2.20 & 159.7 	& 919.7	& 2.32             \\
        \hline
        $\beta-$equilibrium 							& 56.5 & 1005.6 	& 1.48 & 140.1  & 915.1 & 1.73            \\
        \hline
    \end{tabular}
    \caption{Temperature, baryonic chemical potential and baryonic
    density at the CEP for NJL and PNJL. Different scenarios are considered.
    \label{table:CEP} }
\end{center}
\end{table}

In the following we analyze the effect of isospin asymmetry and  we 
center our discussion on the PNJL model. 

{\it Isospin asymmetry.---}In the previous section we have seen that the
location of the CEP depends on isospin. In particular, it was shown
that in  $\beta-$equilibrium matter the CEP occurs at larger baryonic
chemical potentials and smaller temperatures. To study in a
more systematic way the effect of isospin on the CEP we take the
$s$-quark chemical potential equal to zero and increase systematically $\mu_d$
with respect to $\mu_u$. We are interested in the $d$-quark rich matter
as it occurs in HIC and neutron stars: isospin asymmetry presently attained 
in HIC corresponds to $\mu_u<\mu_d< 1.1 \mu_u$, and neutron matter has 
$\mu_d \sim 1.2 \mu_u$. 
Larger isospin asymmetries are possible in $\pi^-$ rich matter 
\cite{cavagnoli11,abuki13}.

In Fig. \ref{dfase2} the results for the CEP obtained for the set 
($\mu_d$, $\mu_u$, $\mu_s=0$) are shown.  
The red full point corresponds to the CEP with $\mu_u=\mu_d=\mu_s$. 
All other CEPs are calculated at $\mu_s=0$, and they all occur for $\rho_s=0$. 
The corresponding densities ($\rho_u,\, \rho_d,\, \rho_s$) are given in 
Table \ref{table:CEP_dens}.
Increasing the isospin asymmetry moves the CEP to smaller temperatures 
and larger baryonic chemical potentials (it can be understood with the 
same reasons as previously for $\beta-$equilibrium case). Eventually, for 
an asymmetry large enough the CEP  disappears. The threshold corresponds 
to $\mu_d=1.45\mu_u$ and is represented in the graph by a star at $T=0$. 
This scenario corresponds to $|\mu_u-\mu_d|=|\mu_I|=|\mu_Q|=130$ MeV, 
below the pion mass and, therefore, no pion condensation occurs
under these conditions. The effect of pion condensation on the QCD
phase diagram for finite chemical potentials has recently been
discussed in \cite{sasaki10,abuki13}. We also remark that in
\cite{ueda13}, where the effect of isospin on the QCD phase diagram
has also  been discussed, a larger isospin chemical potential
corresponds to smaller baryonic chemical potential due to the
definition of the baryonic chemical potential: in \cite{ueda13}, the
study was performed within the SU(2) quark-meson model and the relation
$\mu_B=3 \mu_q=\frac{3}{2}(\mu_u+\mu_d)$ was used; in the present work we
get from  Eq. (\ref{chem}) $\mu_B=\mu_u+2\mu_d$. In both works the temperature 
of the CEP decreases when the isospin asymmetry increases.

\begin{table}[t]
\begin{center}
    \begin{tabular}{|c||c|c|c|c|c|}
        \hline
        CEP	& $T$ [MeV]   & $\mu_B$ [MeV]	& $\rho_B/\rho_0$ & $\rho_u/\rho_0$ & $\rho_d/\rho_0$ \\
        \hline\hline
        $\mu_d=\mu_u$     & 157.5 & 890.4 & 1.74 						& 1.50 						& 1.50 \\
        \hline
        $\mu_d=1.1\mu_u$ 	& 155 & 906 & 1.72 & 1.28 & 1.72 \\
        \hline
        $\mu_d=1.2\mu_u$ 	& 145 & 948 & 1.67 & 1.05 & 1.95 \\
        \hline
        $\mu_d=1.3\mu_u$ 	& 115 & 1029 & 1.69 & 0.75 & 2.25 \\
        \hline
        $\mu_d=1.4\mu_u$ 	& 62 & 1102 & 1.85 & 0.50 & 2.50 \\
        \hline
        $\mu_d=1.45\mu_u$ & $\sim 0$ & 1126 & 1.91 & 0.39 & 2.61 \\
        \hline
    \end{tabular}
    \caption{\label{table:CEP_dens} The temperature, baryonic  chemical 
    potential, and $u(d)$ quark densities at the CEPs for different
    scenarios ratio $\mu_d/\mu_d$ with $\mu_s = 0$ ($\rho_s = 0$). The
    baryonic density is given in terms of the saturation density
    $\rho_0=0.16$ fm$^{-3}$.}
\end{center}
\end{table}

\begin{figure*}[tb]
	\includegraphics[width=0.475\linewidth,angle=0]{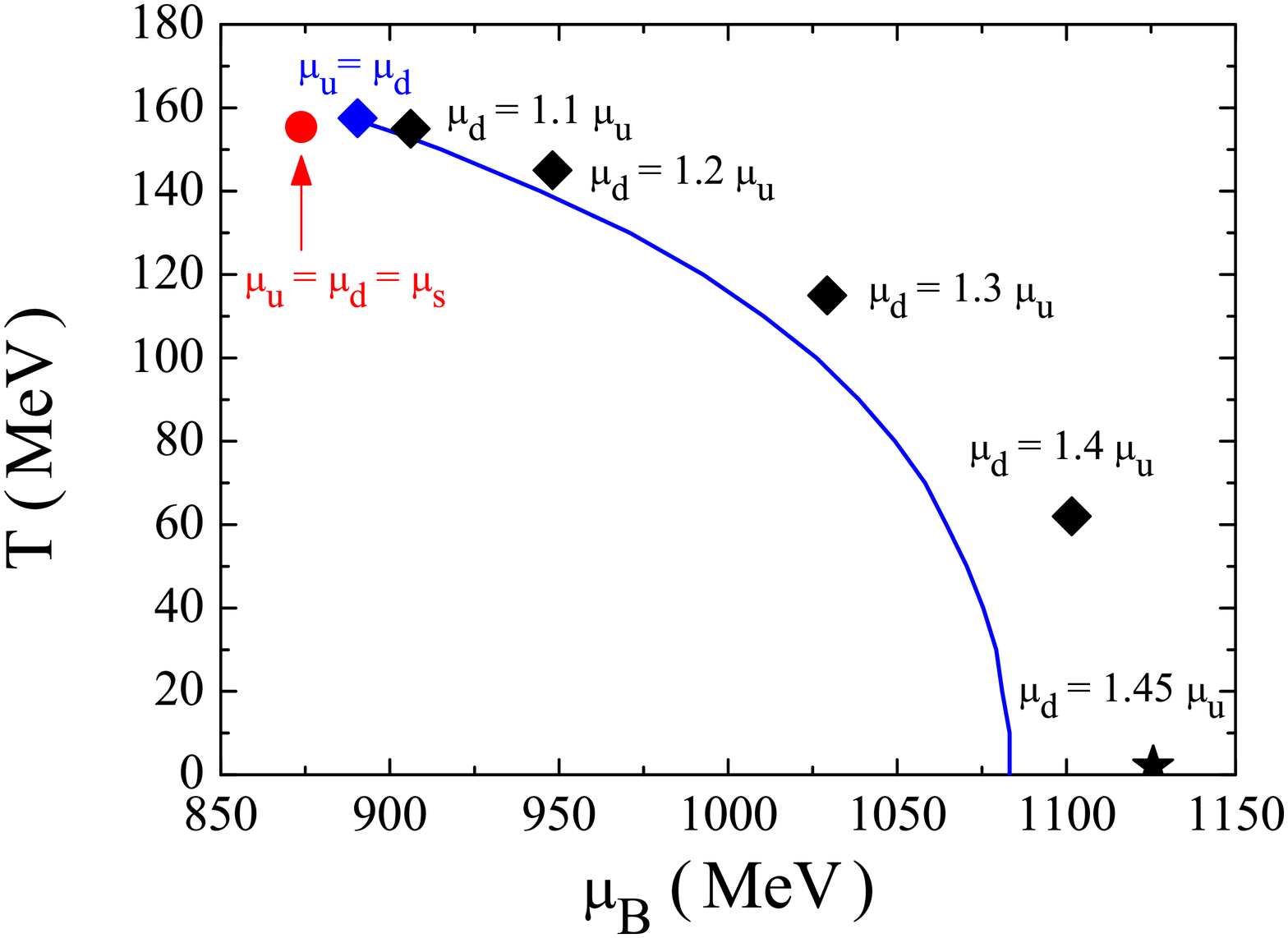}		   		  	 
	\includegraphics[width=0.475\linewidth,angle=0]{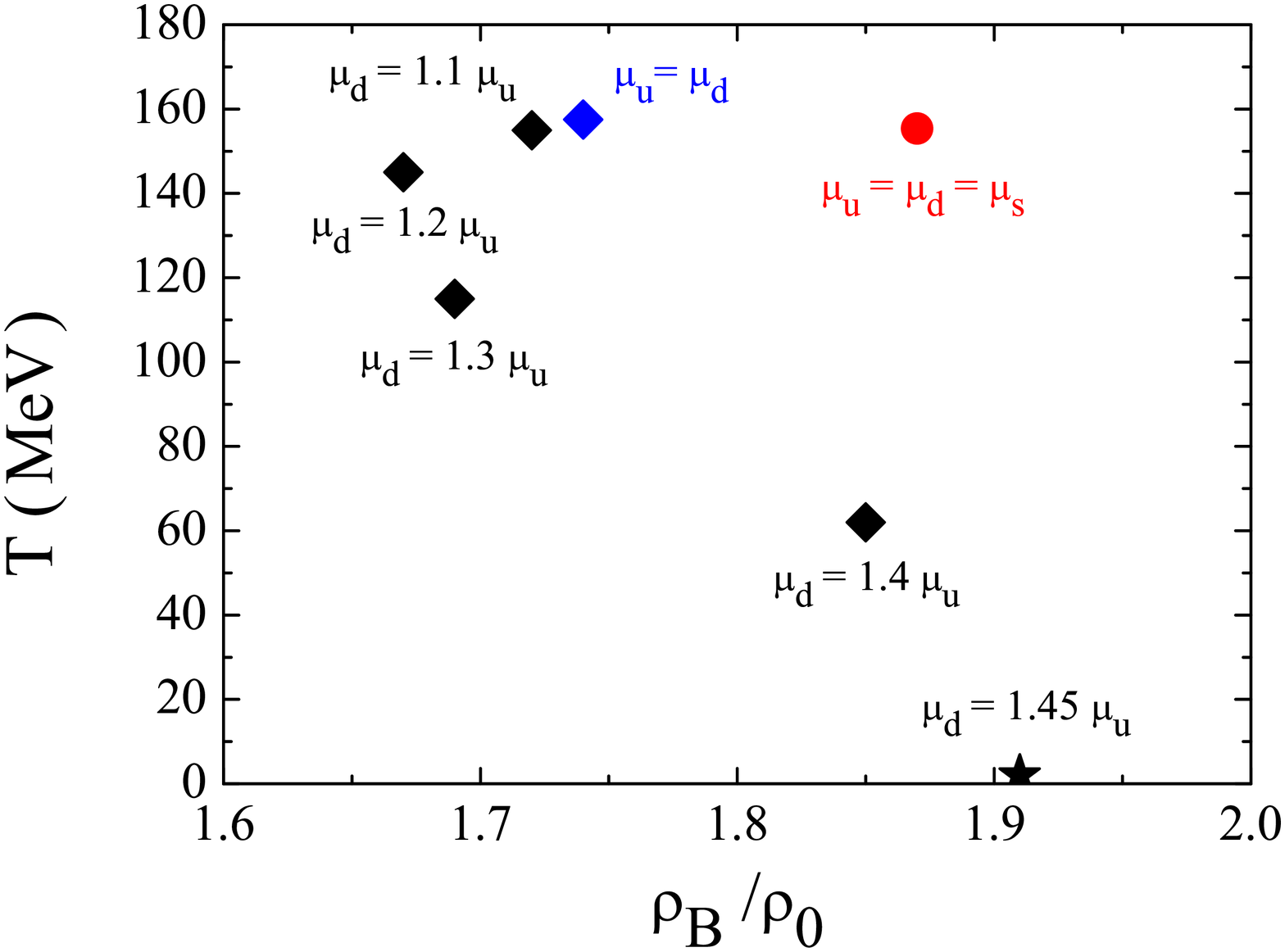}
	\caption{Effect of isospin in the location of the CEP within the PNJL model.  
	The full line is the first order phase transition line for zero isospin matter 
	($\mu_d=\mu_u$, $\mu_s=0$). The strangeness chemical potential was always
	taken equal to zero. For $\mu_d>1.45 \mu_u$  no CEP exists. Isospin asymmetry 
	presently attained in HIC corresponds to $\mu_u<\mu_d< 1.1 \mu_u$ and neutron 
	matter to $\mu_d \sim 1.2 \mu_u$.}
\label{dfase2}
\end{figure*}

In the left panel of Fig. \ref{dfase2}, the CEP is shown for $T$ versus the 
baryonic density. For $\mu_u<\mu_d< 1.2 \mu_u$ the baryonic density of the CEP 
decreases with asymmetry but for $\mu_d \gtrsim 1.2 \mu_u$ the opposite occurs 
and at the threshold ($\mu_d=1.45\mu_u$) $\rho_B\sim 1.91 \rho_0$;
see Table \ref{table:CEP_dens}. 

{\it External magnetic field.---}In the following we study the effect of a static 
external magnetic field on the localization of the CEPs \cite{Scoccola} previously 
calculated and plot the results in Fig. \ref{dfase3}. The values for the CEPs are
given in Table \ref{table:CEP_eB}. The red dots correspond to symmetric
matter with $\mu_u=\mu_d=\mu_s$ and reproduce qualitatively the results 
previously obtained within the NJL  in \cite{avancini2012} to the PNJL model.
The trend is very similar: as the intensity of the magnetic field  increases, 
the transition temperature increases and the baryonic chemical potential 
decreases until the critical value $eB\sim 0.4$ GeV$^2$. For stronger magnetic 
fields both $T$ and $\mu_B$ increase. In the middle panel of Fig. \ref{dfase3} the
CEP is given in a $T$  versus baryonic density plot. It is seen that when $eB$ 
increases from 0 to 1 GeV$^2$ the baryonic density at the CEP increases from 
2$\rho_0$ to $14\rho_0$. 

Taking the isospin symmetric matter scenario $\mu_u=\mu_d$ and
$\mu_s=0$, the effect of the magnetic on the CEP is very similar to
the previous one (see blue triangles in Fig. \ref{dfase3}): $T$
is only slightly larger and the CEP baryonic density slightly smaller.

\begin{figure}[t]
\begin{tabular}{c}
    \includegraphics[width=0.8\linewidth,angle=0]{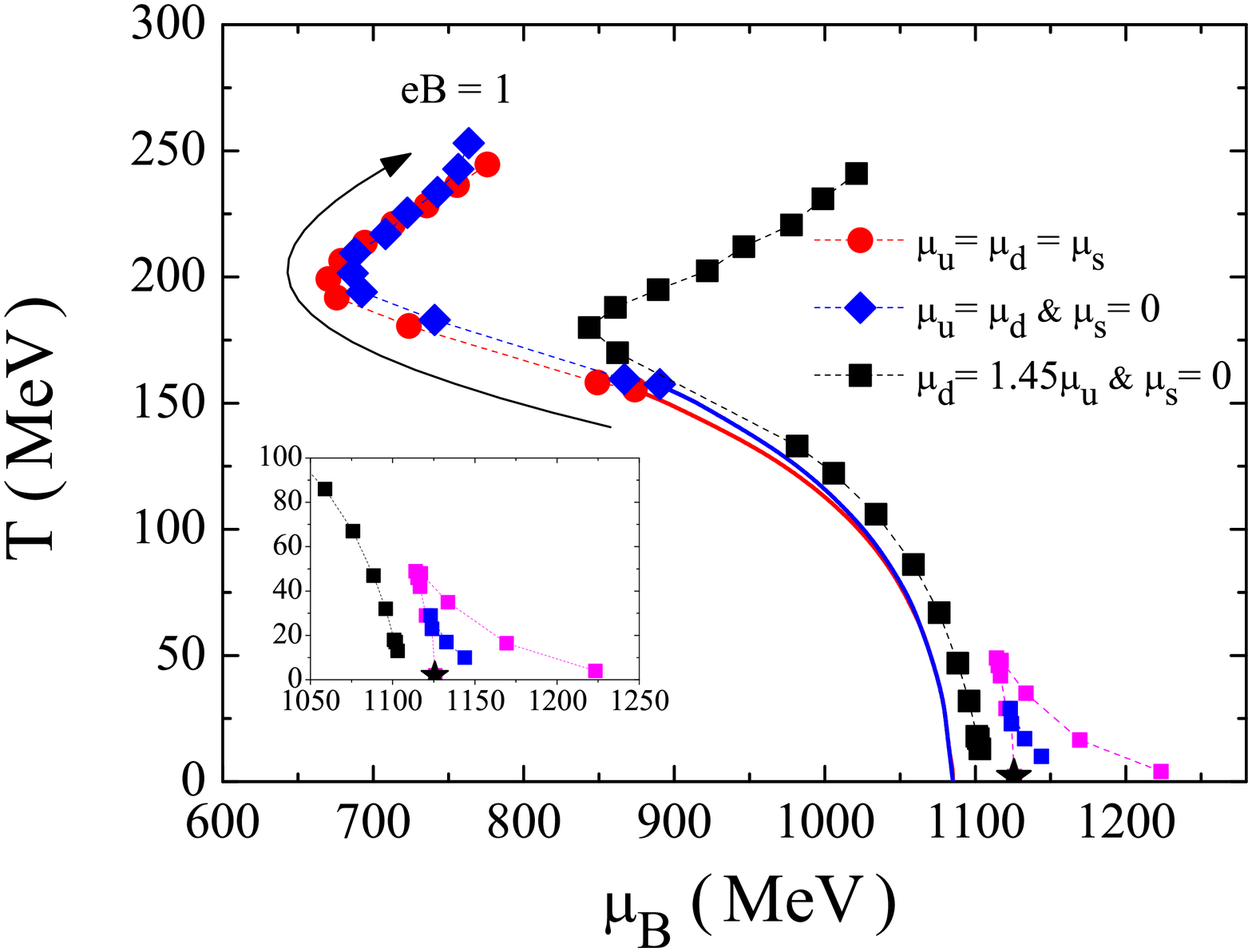}\\
    \vspace{-0.3cm}
    \includegraphics[width=0.8\linewidth,angle=0]{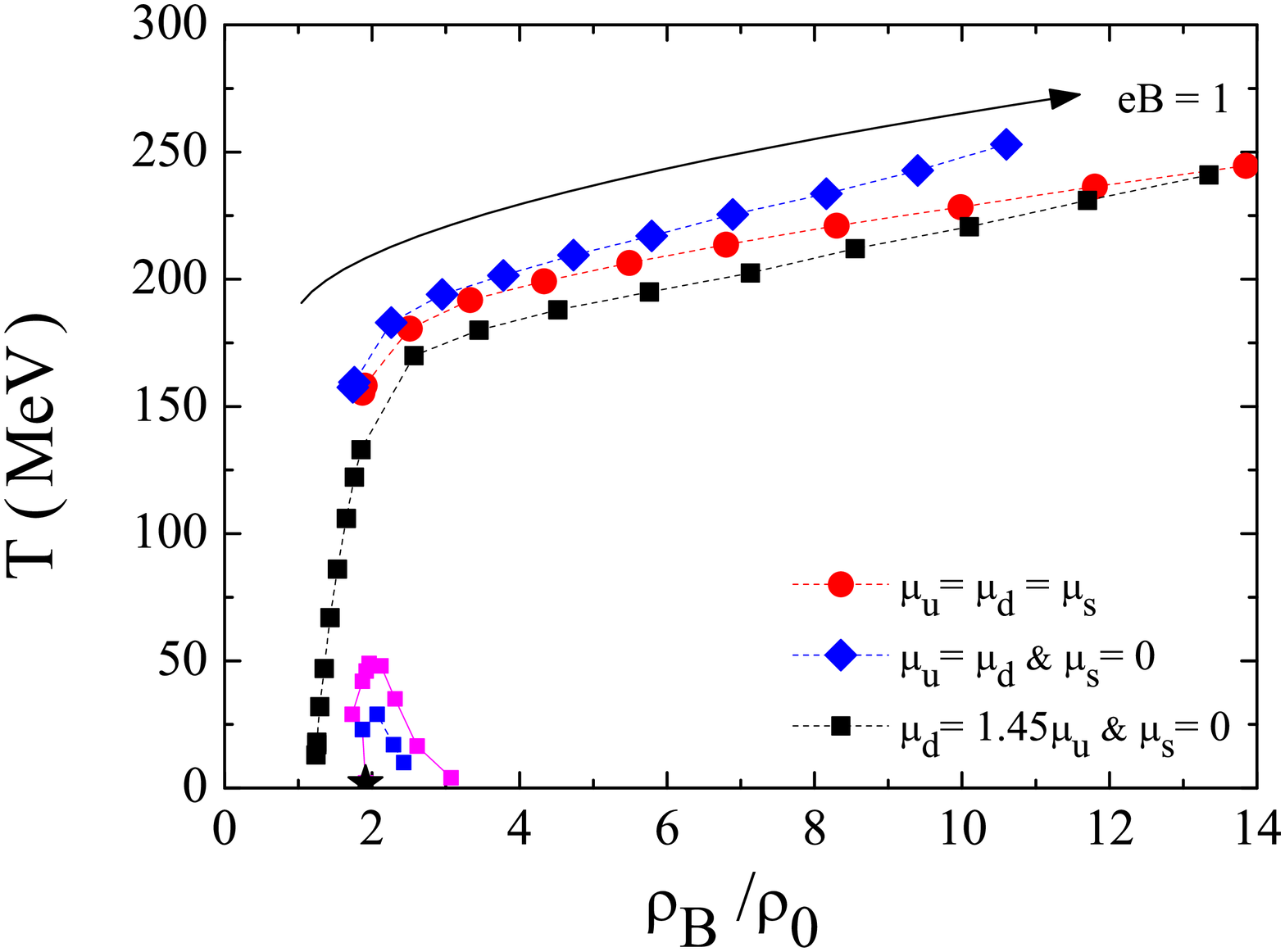}\\
    \vspace{-0.2cm}
    \includegraphics[width=0.8\linewidth,angle=0]{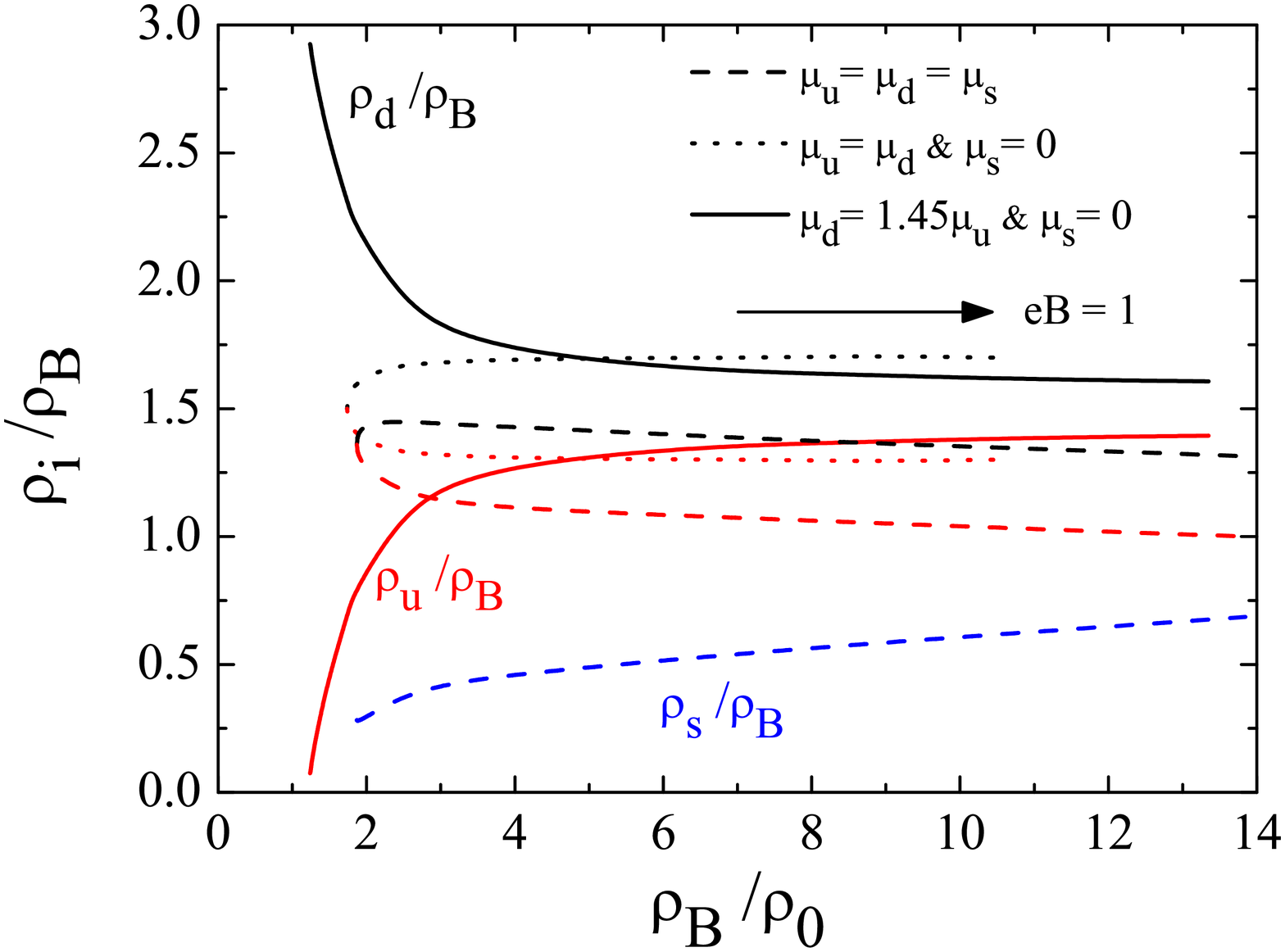}    
\end{tabular}
  \caption{
  Effect of an external magnetic field on the CEP's location within the PNJL model: 
	$T^{CEP}$ versus baryonic chemical potential (top panel) and baryonic density (middle panel).
	The full lines are the first order transitions at $eB=0$.
	Three scenarios are shown: $\mu_u=\mu_d=\mu_s$ (red dots), $\mu_u=\mu_d;\,\mu_s=0$ 
	(blue diamonds) and $\mu_d=1.45 \mu_u$, $\mu_s=0$ (black squares) 
	corresponding to the threshold isospin asymmetry above which no CEP occurs. 
	In the last case for strong enough magnetic fields and low temperatures
	two or more CEPs exist at different temperatures for a given magnetic field 
	intensity (pink and blue squares). The bottom panel shows the $u$, $d$ and $s$ 
	quark fractions as a function of the baryonic density: dashed line corresponds 
	to $\mu_u=\mu_d=\mu_s$, dotted line corresponds to $\mu_u=\mu_d;\,\mu_s=0$ and 
	full line corresponds to $\mu_d=1.45\mu_u;\,\mu_s=0$.
	}
\label{dfase3}
\end{figure}

A very interesting case occurs for the very asymmetric matter scenario: 
a first order phase transition driven by the magnetic field takes place
if $\mu_d\gtrsim1.45 \mu_u$. Taking the threshold value $\mu_d=1.45\mu_u$ it
is seen that for $eB<$0.1 GeV$^2$ two CEPs may appear. 
In fact, for sufficiently small values of $eB$ the $T^{CEP}$ is small and the 
Landau level effects are visible. 

A magnetic field affects in a different way $u$ and $d$ quarks due to their 
different electric charges. A consequence is the possible appearance of
two or more CEPs for a given magnetic field intensity. 
Two critical end points occur at different values of $T$ and $\mu_B$ 
for the same magnetic field intensity for fields $0.03 \lesssim eB
\lesssim 0.07$ GeV$^2$.  Above 0.07 GeV$^2$ only one CEP remains. 
For stronger fields we get  $T^{CEP}>100$ MeV: Landau level effects are 
completely washed out at these temperatures.  In the lower panel of 
Fig. \ref{dfase3} we plot the $u$ and $d$ quark fractions corresponding 
to each CEP at different magnetic fields and for $\mu_d=1.45 \mu_u$ 
versus the baryonic density: it is seen that as the magnetic field 
becomes more intense the fraction of $u$ quarks comes closer to the 
$d$ quark fraction. This is due to the larger charge of the $u$ quarks 
and the fact that the quark density is proportional to the absolute 
value of the charge times the magnetic field intensity.

\begin{table}[t]
\begin{center}
    \begin{tabular}{|c||c|c|c||c|c|c|}
            \hline
            \multicolumn{0}{|c||}{$ $}
                & \multicolumn{3}{|c||}{$\mu_u=\mu_d=\mu_s$}
                & \multicolumn{3}{|c|}{$\mu_u=\mu_d\textmd{; }\mu_s=0$} \\
            \hline
        $eB$    & $T$ & $\mu_B$ & $\rho_B/\rho_0$ & $T$ & $\mu_B$ & $\rho_B/\rho_0$ \\
        $[\mbox{GeV}^2]$    & [MeV]     & [MeV]     & { } & [MeV] & [MeV] &   { }   \\
        \hline\hline
        0   & 155.4 & 873.8	 & 1.87 & 157.5  & 890.4 &  1.74             \\
        \hline
        0.1 & 158.2 & 848.9	 & 1.90 & 159.5  & 866.9 &  1.75             \\
        \hline
        0.2 & 180.6 & 723.8 & 2.51 & 182.8  & 740.8 &  2.25             \\
        \hline
        0.3 & 191.8 & 675.7 & 3.33 & 194.1  & 691.5 &  3.00            \\
        \hline
        0.4 & 199.2 & 670.2 & 4.33 & 201.6  & 686.4 &  3.80         \\
        \hline
        0.5 & 206.4 & 678.6 & 5.49 & 210.0  & 688.0 &  4.72         \\
        \hline
        0.6 & 213.6 & 694.5 & 6.80 & 217.0  & 708.1 &  5.79         \\
        \hline
        0.7 & 221.0 & 713.3 & 8.30 & 225.5  & 722.7 &  6.89         \\
        \hline
        0.8 & 228.4 & 735.5 & 9.98 & 233.6  & 742.7 &  8.10     \\
        \hline
        0.9 & 236.4 & 755.6 & 11.80 & 242.8  & 756.5 & 9.40     \\
        \hline
        1 & 244.6 & 775.9 & 13.85 & 253.0  & 763.5 &  10.6     \\
        \hline
    \end{tabular}
    \caption{\label{table:CEP_eB} The temperature, baryonic  chemical 
    potential and density at the CEPs for different
    values of the magnetic field and two different scenarios:
    $\mu_u=\mu_d=\mu_s$ and $\mu_u=\mu_d\textmd{; }\mu_s=0$. The
    baryonic density is given in terms of the saturation density
    $\rho_0=0.16$ fm$^{-3}$.}
\end{center}
\end{table}

The right panel of Fig. 3 shows the $d$ (black), $u$ (red) and $s$ (blue) 
quark fractions as a function of the baryonic density at the CEPs for the 
three scenarios considered. When $\mu_u=\mu_d=\mu_s$ there is a strange 
quark fraction in the CEP which increases with the baryonic density. 
For the other two scenarios, the $u$ and $d$ quark fractions show a 
tendency to stabilize around $1.5\rho_B$.

Finally, it is also important to point out that all three scenarios presented in 
Fig. \ref{dfase3} show an inverse magnetic catalysis at finite chemical potential 
and zero temperature once the critical temperature decreases with increasing 
$eB$ \cite{Andersen}.
However, at large values of $eB$ the inverse magnetic catalysis tendency disappears 
and a magnetic catalysis takes place.

\section{Conclusions}

In the present study the location of the CEP on the QCD phase diagram
was calculated within different scenarios in the framework of the
SU(3) PNJL model. For reference some results obtained within the 
NJL model have also been shown. 

Different scenarios have been considered, namely with respect to the
isospin and strangeness content of matter.  It was shown that for
$\beta$-equilibrium matter the CEP occurs at smaller temperatures and
densities. This scenario is of interest for neutron stars. However,
the $T^{CEP}$ calculated within PNJL seems too high to occur in a
protoneutron star. These results, however, confirm previous calculations
that indicate that a deconfinement phase transition in the laboratory
will be more easily attained with asymmetric nuclear matter
\cite{ditoro,cavagnoli11}. It was shown that for very asymmetric
matter,  in particular for $\mu_d>1.4\mu_u$, no first order phase
transition to a deconfined phase occurs. The disappearance of the CEP
above a critical  isospin chemical potential was also obtained in
\cite{abuki13} where a Ginzburg-Landau approach was used to study the
QCD phase structure.

We have next studied the effect of strong magnetic fields on the
location of the CEP, generalizing the results of \cite{avancini2012}
to new nonsymmetric scenarios. For  a zero $s$-quark  and null isospin
chemical potential, results very similar to the equal chemical
potentials case were obtained. A more interesting situation was
observed when analyzing very isospin asymmetric matter: in this case
starting from a scenario having an isospin asymmetry above which the
CEP does not exist for a zero external magnetic field it was shown
that a sufficiently high  external magnetic field could drive the
system to a first order phase transition. The critical end point
occurs at very small temperatures if $eB<0.1$ GeV$^2$ and, in this
case, a complicated structure with several CEP at different values of
($T, \mu_B)$ is possible for the same magnetic field, because the
temperature is not high enough to wash out the Landau level
effects. For $eB>0.1$ GeV$^2$ only one CEP exists.

This is an important result because it shows that a strong magnetic
field is able to drive a system with no CEP into a first order phase
transition. In the present study we have explored the possibility that
this occurs in a very isospin asymmetric system. 
A quite different situation could also give rise to a similar result: 
it has been shown that including  a vector repulsive term in the quark 
model Lagrangian density it is possible to obtain a phase diagram with 
no first order chiral phase transition\cite{weise,ueda13}. 
We may expect that a  strong enough field would drive the system into 
a first order phase transition. 
This behavior would go along the results obtained in \cite{marcus}.

It has been shown that  at zero baryonic chemical potential lattice QCD 
calculations predict a decreasing deconfinement critical temperature with 
an increase of $eB$ \cite{baliJHEP2012}. This behavior  is not obtained 
within PNJL model \cite{Ferreira:2013tba}. Therefore, in order to confirm the 
present results it is important to include possible  back reaction effects 
of the external magnetic field  on the Polyakov loop.

\vspace{0.5cm}
{\bf Ackowledgements}: 
This work was partially supported by Projects No. PTDC/FIS/113292/2009 and No. 
CERN/FP/123620/2011 developed under the initiative QREN financed by the UE/FEDER 
through the program COMPETE $--$ ``Programa Operacional Factores de 
Competitividade'', by Grant No. SFRH/BD/51717/2011, by CNPq/Brazil and FAPESC/Brazil.


\end{document}